%% file: Migdal_NTDT3.tex
\documentclass[utf8]{frontiersinFPHY_FAMS} % Vancouver Reference Style (Numbered) for Frontiers in Physics

\setcitestyle{square}
\usepackage{url,hyperref,lineno,microtype,subcaption}
\usepackage[onehalfspacing]{setspace}
\usepackage{mathtools}
\hypersetup{colorlinks=true,linkcolor=blue,citecolor=blue,urlcolor=blue}

\newcommand{\figslot}[2]{\IfFileExists{#1}{\includegraphics[width=#2]{#1}}{\fbox{\parbox{#2}{\centering\vspace{2cm}Figure file \texttt{\detokenize{#1}} to be added.\vspace{2cm}}}}}

\theoremstyle{remark}

\newcommand{\vect}[1]{\boldsymbol{#1}}
\def\NS{\NS{}}

\def\VEV#1{\left\langle #1 \right\rangle}
\def\I{\mathrm{i}}

\def\lrb#1{\left(#1\right)}

\input{mymacros}

\def\keyFont{\fontsize{8}{11}\helveticabold }
\def\firstAuthorLast{Migdal and Wang}
\def\Authors{Alexander Migdal\,$^{1,*}$ and Wei Wang\,$^{2}$}
\def\Address{$^{1}$Institute for Advanced Study, Princeton, NJ, USA \\
$^{2}$Department of Mechanical and Aerospace Engineering, New York University, Brooklyn, NY 11201, USA}
\def\corrAuthor{Alexander Migdal\\Institute for Advanced Study, 1 Einstein Drive, Princeton, NJ 08540, USA}
\def\corrEmail{amigdal@ias.edu}

\begin{document}
\onecolumn
\firstpage{1}

\title[Number Theory of Decaying Turbulence 3]{Number Theory of Decaying Turbulence 3: Comparing with Physical and Numerical Experiments}

\author[\firstAuthorLast ]{\Authors}
\address{}
\correspondance{}
\extraAuth{}

\maketitle

\begin{abstract}
\section{}
We compare the predictions of the Euler-ensemble solution of freely decaying incompressible turbulence with simulations
and experiments: the scaling function of the energy spectrum, the index of the second-order structure function and the
decay laws. The predictions have a fixed shape; only their normalization and scale are fitted. In the final \(4096^3\)
direct numerical simulations of Rodhiya and Sreenivasan, the turbulent attractor is selected by a parabola law for the
growth of the bulk length. Along it the collapsed spectrum depends only on the instantaneous Reynolds number, identically
for Saffman and Loitsyansky initial spectra, and its deviation from the theory decreases as the Reynolds number grows;
extrapolated to infinite Reynolds number, it agrees with the theory to \(1\%\) in the bulk of the enstrophy range, below
the dissipation range of the simulation. The energy balance holds exactly. The energy integral of the theory diverges in
the infrared, and the global integrals, energy and enstrophy, depend on how a finite flow cuts off the spectrum: with a
sharp cutoff \(k_0\), the decay laws energy \(\propto t^{-5/4}\) and enstrophy \(\propto t^{-9/4}\) are reached only when
the cutoff enters the \(k^{-7/2}\) tail of the spectrum. Such a cutoff is only an estimate; the reliable test is the bulk
of the spectrum, where the boundaries are not relevant. Fitted with a sharp cutoff, the simulations place it at the lower
edge of the bulk, where the spectrum still carries the initial data, and the measured decay exponents depend on the
initial infrared spectrum. In the Max Planck wind tunnel at \(Re_\lambda\) up to \(5779\), the longitudinal index
extrapolated to infinite Reynolds number matches the odd ensemble with one scale parameter at large separations, where
standard crossover forms fail. The freely decaying turbulent blob of Matsuzawa \emph{et al.} is too small, too
inhomogeneous and too weak to display the attractor. The theory is derived in the first two papers of this series.

\tiny
\keyFont{\section{Keywords:} decaying turbulence, direct numerical simulation, wind tunnel, energy spectrum, structure function, decay laws, Reynolds-number extrapolation, Euler ensemble}
\end{abstract}

\newpage
\section{Introduction}

This paper is the third of three that form the exact analytical sequel to the invited review~\cite{ReviewPaperAM} on
the loop-space theory of decaying turbulence, published together with the \(4096^3\) DNS study of Rodhiya and
Sreenivasan~\cite{SreeniAkash2025}. The first paper~\cite{NTDT1} derives the operator representation of the decaying
statistical problem, its exact solution, the Euler ensemble, and the stability and dimension independence of this
solution; the second~\cite{NTDT2} computes its energy spectrum and analyzes the Riemann-zeta structure of its asymptotics.
The present paper compares the predictions with direct numerical simulations and experiments. Each paper is
self-contained: results of the other two are quoted without derivation.

\subsection*{The predictions tested}

The statistical evolution of incompressible \NS{} flow is rewritten exactly in terms of the circulation loop functional
and mapped to a dual momentum loop, whose diffusive dynamics has an exact decaying solution, the Euler ensemble of random
walks on regular star polygons~\cite{NTDT1}. Its odd-parity sector is the locally stable turbulent attractor. It
solves the loop equation in the local limit, in which the molecular viscosity vanishes at fixed turbulent viscosity, so
the theory describes decaying turbulence at infinite Reynolds number. Its energy spectrum has the self-similar form
\(E(k,t)\propto\tilde\nu^2L^{-1}H(kL)\), \(L\sim\sqrt{\tilde\nu t}\)~\cite{migdal2024quantum}, with a scaling function \(H\) that does not depend on the
dimension of space and is given by a Mellin--Barnes integral with a ratio of Riemann zeta functions~\cite{NTDT1,NTDT2}.
Its local index varies continuously, without an extended \(k^{-5/3}\) range. The same theory gives the second-order
structure function, whose index falls continuously from \(2\) to \(0\) without an inertial plateau, and the decay laws of
a finite flow at late times, energy \(\propto t^{-5/4}\) and enstrophy \(\propto t^{-9/4}\)~\cite{migdal2024quantum,NTDT2}. These predictions are summarized in
Sec.~\ref{sec:predictions}.

\subsection*{Principles of the comparison}

\begin{enumerate}
    \item \textbf{A fixed shape.} The predictions compared with the data have a fixed shape; only their normalization and
    scale are fitted.
    \item \textbf{The turbulent attractor.} A flow approaches the attractor only after its initial data are forgotten and
    before it decays. We divide the data into three stages: the stochastization stage, the turbulent attractor and decayed
    turbulence, and fit only the attractor. In the DNS the attractor is selected by a parabola law for the growth of the
    bulk length, in the wind tunnel by the collapse of the runs at fixed index.
    \item \textbf{Infinite Reynolds number.} The theory is the limit of infinite Reynolds number, so we extrapolate the
    data, not the theory: in the DNS along the decay, with the instantaneous Reynolds number of each snapshot, and in the
    wind tunnel across runs at different Reynolds numbers.
    \item \textbf{The measured function.} The DNS give the total velocity increment, the experiments its longitudinal
    component; the theory provides both (Sec.~\ref{sec:longitudinal}).
\end{enumerate}

\subsection*{Summary of the results}

In the final \(4096^3\) DNS of freely decaying turbulence~\cite{SreeniAkash2025} (Sec.~\ref{sec:dns_match}), the collapsed
spectrum depends only on the instantaneous Reynolds number, for Saffman and Loitsyansky initial spectra alike, and
approaches the theory as the Reynolds number grows; extrapolated to infinite Reynolds number, it agrees with the theory to
\(1\%\) rms in the bulk of the enstrophy range, below the dissipation range of the simulation. The structure-function
index is the same for Saffman and Loitsyansky initial spectra and follows the theory with systematic deviations of up to
\(0.17\), which we attribute to the finite Reynolds number. The energy balance holds exactly. The decay exponents depend
on the initial infrared spectrum. They come from global integrals, which the theory in infinite space determines only up
to the infrared cutoff of a finite flow; estimated with a sharp cutoff fitted to the data, this cutoff lies at the lower
edge of the bulk of the spectrum, far above the lowest shell of the box (Sec.~\ref{sec:decaylaws}). At much higher Reynolds numbers,
the longitudinal index measured in the Max Planck Variable Density Turbulence Tunnel at
\(Re_\lambda=413\)--\(5779\)~\cite{GregXi2} and extrapolated to infinite Reynolds number agrees with the odd Euler ensemble
with one scale parameter in the turbulent-attractor region, while von K\'arm\'an crossover forms do not
(Sec.~\ref{sec:windtunnel}). The turbulent blob of Matsuzawa \emph{et al.}~\cite{Matsuzawa2026} is too small, too
inhomogeneous and too weak to display the attractor (Sec.~\ref{sec:blob}).

\subsection*{Organization of the paper}

Section~\ref{sec:predictions} summarizes the predictions, including the relation between the longitudinal and the total
structure function. Section~\ref{sec:dns_match} compares them with the DNS, Sec.~\ref{sec:windtunnel} with the Max Planck
wind tunnel, and Sec.~\ref{sec:blob} with grid turbulence and the turbulent blob. Section~\ref{sec:discussion} discusses
the results. Details of the data processing are given in the Supplementary Material.

\section{The predictions}
\label{sec:predictions}

This section collects the results of the companion papers~\cite{NTDT1,NTDT2} that are compared with the data.

\subsection{Energy spectrum}
\label{sec:mellinspectrum}

The Euler ensemble gives the decaying energy spectrum in the self-similar form~\cite{migdal2024quantum}
\[
    E(k,t)\propto\frac{\tilde\nu^2}{L(t)}H(kL(t)),
    \qquad
    L(t)\sim\sqrt{\tilde\nu t}.
\]
Here \(\tilde\nu=\nu N^2\) is the turbulent viscosity of the Euler ensemble, which solves the loop equation in the local
limit \(N\to\infty\) at fixed \(\tilde\nu\), so that the molecular viscosity \(\nu\) vanishes: the theory describes freely
decaying turbulence at infinite Reynolds number~\cite{NTDT1}. Flows at finite Reynolds number approach this state only as
an intermediate asymptotic one; at fixed \(\nu\) a real flow eventually leaves it, in the final period of decay. With
\(\kappa=k\sqrt{\tilde\nu t}\), the scaling function, the same in every dimension \(d>1\)~\cite{NTDT1}, is the inverse
Mellin transform~\cite{NTDT2}
\begin{equation}
    H(\kappa)
    =
    \int_{\epsilon-\I\8}^{\epsilon+\I\8}
    \frac{dp}{2\pi\I}\,
    M_\eta(p)\,\kappa^p ,
    \qquad
    M_\eta(p)
    =
    \frac{
    f(p)\Gamma(-p)\zeta\left(p+\frac{15}{2}\right)
    }{
    (2p+7)(2p+17)\zeta\left(p+\frac{17}{2}\right)\lrb{1-\eta\,2^{-(p+17/2)}}
    },
    \label{MellinInverseH}
\end{equation}
with the contour in the strip \(-7/2<\Re p<0\), \(\eta=1\) for the odd ensemble, the attractor, and \(\eta=0\) for the even
one. The entire function \(f(p)\) is generated by the continuum limit of the Euler-ensemble path integral~\cite{NTDT2}.
The even and odd ensembles give spectra within \(0.004\) of each other, below the resolution of the data. The local
spectral index \(d\log H/d\log\kappa\) decreases continuously from \(0\) to \(-7/2\), the exponent of the asymptotic tail
set by the pole \(p=-7/2\); there is no extended \(k^{-5/3}\) range~\cite{NTDT2}.

\subsection{Structure function}

Multiplying~\eqref{MellinInverseH} by the kernel \(2\left(1-\sin(\kappa x)/(\kappa x)\right)\) and integrating over
\(\kappa\) gives the scaling function of the second-order structure function \(\VEV{(\Delta\vect v)^2}(r,t)\) in the
variable \(x=r/L(t)\), up to normalization~\cite{NTDT2}:
\begin{equation}
    D(x)
    =2\int_0^\infty d\kappa\,H(\kappa)\left(1-\frac{\sin\kappa x}{\kappa x}\right)
    =\int_{c-\I\infty}^{c+\I\infty}\frac{dq}{2\pi\I}\,x^{q}\left(-2Z_\eta(q)\right),
    \qquad 0<c<2,
    \label{eq:Dx}
\end{equation}
with \(Z_\eta(q)=-\Gamma(-1-q)\cos(\pi q/2)\,M_\eta(-1-q)\). Its effective index
\(\zeta_2(x)=x\,\partial_x\log D(x)\), computed by the Lefschetz-thimble method~\cite{NTDT2,MigdalWindTunnel2026}, falls
monotonically from the dissipative value \(2\) at \(x\to0\) to \(0\) at \(x\to\infty\), and it has no inertial plateau. The
whole crossover from \(\zeta_2=2\) to \(\zeta_2\approx0\) takes place within \(-5\lesssim\log x\lesssim0\).

\subsection{Longitudinal against isotropic correlation}
\label{sec:longitudinal}

The function~\eqref{eq:Dx} is the scaling function of the total increment, the trace of the second-order structure
tensor, \(\VEV{(\Delta\vect v)^2}=D_{LL}+2D_{NN}\). Measurements of one velocity component along the separation give the
longitudinal function \(D_{LL}\), and of a component normal to it the transverse function \(D_{NN}\). In isotropic
incompressible turbulence the two are related by \(D_{NN}=D_{LL}+\frac r2\,dD_{LL}/dr\), so that
\(\VEV{(\Delta\vect v)^2}=r^{-2}d(r^3D_{LL})/dr\), and the longitudinal function follows from~\eqref{eq:Dx} as
\begin{equation}
    D_{LL}(x)=x^{-3}\int_0^x ds\,s^2D(s),
    \label{eq:DLL}
\end{equation}
which divides the Mellin amplitude in~\eqref{eq:Dx} by \(q+3\). Isotropy relates the three functions, it does not make
them equal: their indices coincide for a pure power law but not for the running index, and the indices of \(D\) and
\(D_{LL}\) differ by up to \(0.14\) in the crossover. We compare the DNS, whose structure function is computed from the
spectra, with~\eqref{eq:Dx}, and the experiments of Secs.~\ref{sec:windtunnel} and~\ref{sec:blob}, which measure the
longitudinal function, with~\eqref{eq:DLL}.

\subsection{Decay laws}
\label{sec:decaylaws}

The classical decay laws follow from the conservation of an infrared invariant. The Saffman integral gives
\(E\propto t^{-6/5}\) and an integral scale \(\ell\propto t^{2/5}\)~\cite{Saffman1967}, the Loitsyansky integral
\(E\propto t^{-10/7}\) and \(\ell\propto t^{2/7}\)~\cite{DavidsonTurbulence2015,Ishida2006,JohnDonzisSreenivasan2022}.
The Euler ensemble instead predicts the diffusive growth \(L(t)\propto\sqrt{t+t_0}\) and, with the normalization
\(E(k,t)\propto\tilde\nu^2L^{-1}H(kL)\)~\cite{migdal2024quantum}, the enstrophy
\begin{equation}
    En(t)=\int_{k_0}^\infty dk\,k^2E(k,t)\propto\frac{\tilde\nu^2}{L^4}\int_{k_0L}^\infty d\kappa\,\kappa^2H(\kappa),
    \label{eq:enstrophyIntegral}
\end{equation}
where \(k_0\) is the infrared cutoff of the scaling range, \(k_0\sim1/(\text{system size})\) in a finite
flow~\cite{migdal2024quantum}. At late times, once \(k_0L(t)\) lies in the tail \(H\propto\kappa^{-7/2}\), the integral
is dominated by its lower limit,
\begin{equation}
    En(t)\sim\frac{1}{L^4}\int_{k_0L}^\infty d\kappa\,\kappa^2\kappa^{-7/2}\propto\frac{1}{L^{9/2}}\propto t^{-9/4},
    \label{eq:enstrophyDecay}
\end{equation}
and the energy decays as \(E\propto t^{-5/4}\)~\cite{migdal2024quantum,ReviewPaperAM,NTDT2}. In every freely decaying
homogeneous flow the two are tied by the energy balance \(-dE/dt=2\nu En\), so that \(E\propto t^{-n}\) requires
\(En\propto t^{-n-1}\): the energy and the enstrophy cannot follow different decay laws.

In an infinite flow only the enstrophy integral is determined by the scaling function. It converges at both ends, since \(H(\kappa)\) is
finite at \(\kappa=0\) and \(\kappa^2H\propto\kappa^{-3/2}\) at large \(\kappa\). The energy integral diverges in the
infrared. The loop equation fixes the vorticity correlations, and the velocity only up to a potential part, the gradient
of a harmonic function, which carries no vorticity and in an incompressible flow lives at \(k\to0\), at the scale of the
system. In this theory the energy has no finite statistical limit~\cite{migdal2024quantum}. The decay laws refer to
\(E(t)-E(\infty)\), where the constant \(E(\infty)\) is the residual energy of the laminar flow that remains after the
rotational part has dissipated, and the decaying part follows from the dissipation,
\(E(t)-E(\infty)=\int_t^\infty dt'\,2\nu En(t')\)~\cite{migdal2024quantum}. The total energy of a finite flow is carried
by the smallest wave numbers, where the spectrum is set by the boundaries~\cite{migdal2024quantum} and by the initial
data (Sec.~\ref{sec:dns_decay}). In a finite flow at late times the enstrophy integral, too, is dominated by its lower
limit~\eqref{eq:enstrophyDecay}, and its value then depends on how the flow cuts off the spectrum. A physical system
introduces a cutoff function rather than the sharp cutoff of~\eqref{eq:enstrophyIntegral}, and an integral dominated by a
power-law end depends on its form. The sharp cutoff is therefore only an estimate: the global integrals, energy and
enstrophy, cannot be computed reliably from the theory in infinite space. The reliable test is the bulk of the spectrum,
where the boundaries are not relevant, and this is what we compare with the data.

The decay exponents depend on \(k_0L(t)\). In an infinite flow, \(k_0=0\), the integral in~\eqref{eq:enstrophyIntegral}
is a constant, \(En\propto t^{-2}\) and \(E-E(\infty)\propto t^{-1}\)~\cite{migdal2024quantum}. The scaling function
reaches its tail only at \(\kappa\approx10^2\): its local index is \(-1.8\) at \(\kappa=55\), \(-3.2\) at \(\kappa=150\)
and \(-3.48\) at \(\kappa=400\). Evaluated with the sharp cutoff of~\eqref{eq:enstrophyIntegral}, the exponents \((n_E,n_{En})\) of the energy
and the enstrophy are \((1,2)\) for \(k_0L\lesssim10\), \((1.16,2.10)\) at \(k_0L=55\) and \((1.24,2.23)\) at
\(k_0L=150\), and they reach the asymptotic values \((5/4,9/4)\) only for \(k_0L\gtrsim400\)
(Fig.~\ref{fig:DNSdecay}a). In the length unit of the spectral fit, \(L=e^{\xi_0}L_n\) with \(e^{\xi_0}\approx95\)
(Sec.~\ref{sec:dns_collapse}), the asymptotic laws need \(k_0L_n\gtrsim1\)--\(4\), a cutoff at the bulk of the spectrum.
The cutoff is not known a priori; in Sec.~\ref{sec:dns_decay} we estimate it by fitting~\eqref{eq:enstrophyIntegral} to the
simulations.

\section{Direct numerical simulation}
\label{sec:dns_match}

\subsection{DNS collapse and direct spectral comparison}
\label{sec:dns_collapse}

\begin{figure}[htbp]
    \centering
    \includegraphics[width=\linewidth]{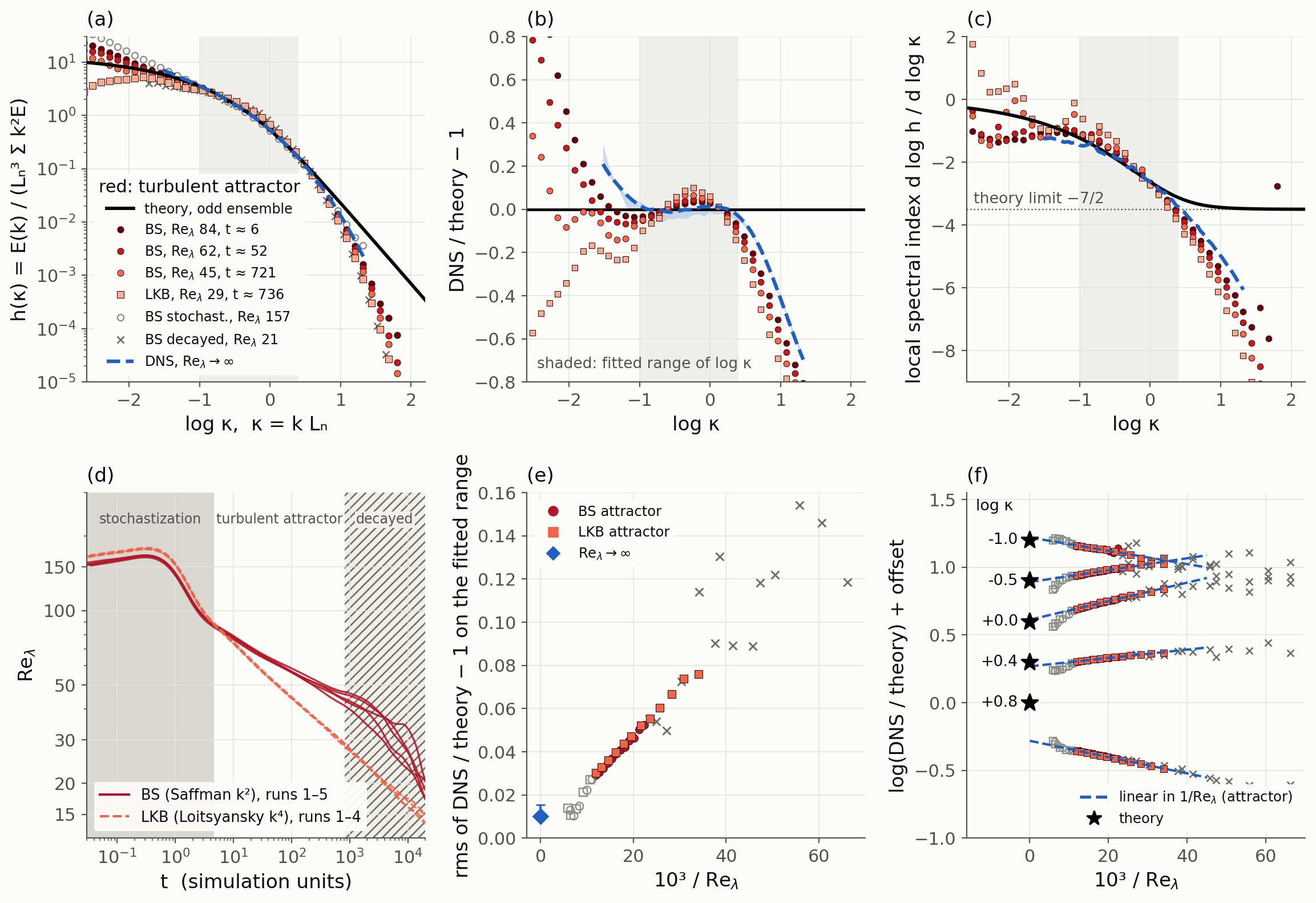}
    \caption{Scaling function of the \(4096^3\) DNS~\cite{SreeniAkash2025} against the odd Euler ensemble, over the
    whole decay. Red: the turbulent attractor, \(4.6\le t\le817\), selected by the parabola law~\eqref{eq:parabolaLaw},
    the only data fitted; open grey: the stochastization stage; grey crosses: decayed turbulence. Circles: BS runs;
    squares: LKB runs; darker red: higher \(Re_\lambda\). Black: the theory \(\mathcal A H(\log\kappa+\xi_0)\) from the
    Mellin--Barnes integral~\eqref{MellinInverseH}. Blue dashed: the DNS extrapolated to \(Re_\lambda\to\infty\), with the
    spread of the extrapolation variants (band), drawn where the spread is below \(0.1\) in \(\log h\). The shaded range
    of \(\log\kappa\) is fitted. (a) The collapsed spectrum~\eqref{DNScollapse}, \(\kappa=kL_n\), averaged over time bins
    and runs, at four Reynolds numbers of the attractor and at the earliest and the latest times; the theory carries the
    constants fitted to the extrapolated spectrum. (b) DNS/theory\(\,-1\), each curve with its own \(\mathcal A\) and
    \(\xi_0\). (c) Local spectral index; the theory tends to \(-7/2\). (d) \(Re_\lambda(t)\) of the nine runs and the
    three stages. (e) The rms deviation over the fitted range in every time bin against \(1/Re_\lambda\); the bar on the
    extrapolated value is the spread of the variants. (f) \(\log(h/h_{\rm th})\) at fixed \(\log\kappa\) against
    \(1/Re_\lambda\), offset for clarity, with the constants of the extrapolated spectrum; lines: linear fits through the
    attractor bins; stars: the theory. The run-to-run errors are smaller than the symbols.}
    \label{fig:TheoryVsDNSCompare}
\end{figure}

We compare the theory with the \(4096^3\) direct numerical simulation of freely
decaying homogeneous isotropic turbulence by Rodhiya and Sreenivasan~\cite{SreeniAkash2025}, in its final form: five
runs with a Birkhoff--Saffman (BS) and four with a Loitsyansky--Kolmogorov--Batchelor (LKB) initial spectrum
(Sec.~\ref{sec:dns_decay}), each with \(736\)--\(2039\) snapshots of the energy spectrum \(E(k,t_n)\), unequally spaced in
time up to \(t\approx10^5\). The runs of each family start from the same shell spectrum with independent phases. The
Taylor Reynolds number \(Re_\lambda=u'\lambda/\nu\), with \(u'^2=\frac23\sum_kE\) and \(\lambda^2=5\sum_kE/\sum_kk^2E\),
is about \(150\) (BS) and \(165\) (LKB) at the start (nominally \(145\)~\cite{SreeniAkash2025}) and decays with time
(Fig.~\ref{fig:TheoryVsDNSCompare}d). The data do not record the viscosity: we take \(\nu\) from the energy balance,
which holds to \(1\%\) in every run (Sec.~\ref{sec:dns_decay}), and normalize \(Re_\lambda\) so that the dissipation
range lies at its universal position in \(k\eta\); the velocity-derivative skewness from the enstrophy budget is then
\(S\approx-0.5\), the standard value at these Reynolds numbers. One BS run (run~4) stores its spectra with an extra
overall factor, exposed by its skewness, while its dynamics coincide with those of the other BS runs; the collapsed
spectra below do not depend on the normalization of \(E\), and the extrapolation to \(1/Re_\lambda\to0\) does not depend
on that of \(Re_\lambda\) (Supplementary Material, Sec.~S1.1).

For each snapshot we define the empirical bulk length scale
\begin{equation}
    L_n
    =
    \exp{
    -
    \frac{
    \sum_k E(k,t_n)k^2\log k
    }{
    \sum_k E(k,t_n)k^2
    }
    } ,
    \label{DNSLengthScale}
\end{equation}
the rescaled variable \(\kappa=kL_n\) and the scale-free collapsed spectrum
\begin{equation}
    h_n(\kappa)=\frac{E(\kappa/L_n,t_n)}{L_n^3\sum_k k^2E(k,t_n)} .
    \label{DNScollapse}
\end{equation}
This enstrophy-weighted logarithmic mean suppresses infrared finite-box effects and fixes the empirical length gauge.
For a self-similar spectrum \(E=c\,L^{-a}H(kL)\), \(h_n\) does not depend on the snapshot, whatever \(c\) and \(a\); without
the factor \(L_n^{-3}\) it would grow as \(L_n^3\). Each snapshot's \(h_n\) is interpolated in \(\log\kappa\) between its own shells, so that every snapshot carries equal
weight at every \(\kappa\); the collapsed spectra are averaged over the snapshots in bins of time a factor \(1.56\) wide,
first in each run and then over the runs, and each bin carries its mean \(1/Re_\lambda\). The error bars are the
run-to-run standard errors, since successive snapshots are strongly correlated.

\paragraph*{Choice of the length scale.}
The length used to collapse the data must be proportional to \(L(t)\) on the attractor. For a self-similar spectrum every
ratio of moments \(\int k^aE\,dk/\int k^bE\,dk\) is \(L(t)^{b-a}\) times a constant, as long as both integrals are dominated
by the self-similar bulk, which favours the enstrophy-weighted moments, \(a,b\ge1\). The length~\eqref{DNSLengthScale} is
the enstrophy-weighted geometric mean of \(1/k\); the length \(L_M=\int kE\,dk/\int k^2E\,dk\) of
Refs.~\cite{ReviewPaperAM,SreeniAkash2025} is the corresponding arithmetic mean. The integral scale
\(\ell=(3\pi/4)\int k^{-1}E\,dk/\int E\,dk\) instead weights the smallest wave numbers, where the initial form of the
spectrum, the finite box and the forcing persist (Secs.~\ref{sec:dns_decay} and~\ref{sec:windtunnel}). The
length~\eqref{DNSLengthScale} is also the normalization \(M'(-3)=0\) of each snapshot's Mellin transform at the enstrophy
moment \(p=-3\).

\paragraph*{Turbulent attractor.}
As in Ref.~\cite{ReviewPaperAM}, the snapshots on the turbulent attractor are selected by the parabola law
\begin{equation}
    t_n=a+b\,L_n+c\,L_n^2 ,
    \label{eq:parabolaLaw}
\end{equation}
that is, \(L_n=L_0+\sqrt{(t_n-t_0)/c}\): the diffusive growth \(L\propto\sqrt t\) of Sec.~\ref{sec:mellinspectrum},
with offsets in time and length. Fitted with weights \(1/t_n\), the law holds in each of the nine runs from \(t\approx3\)
to \(t\approx10^3\), with \(c^{-1}=0.49\)--\(0.50\,\nu\) in the BS runs and \(0.60\,\nu\) in the LKB runs. Over \(4.6\le t\le817\),
the snapshots \(N=120\)--\(1000\) of BS run~1 and the same times in the other runs, its rms deviation is \(0.7\)--\(1.2\%\)
of \(t\). Starting the window at \(t=2.2\) raises the maximum deviation to \(9\)--\(13\%\); extending it to \(t=3050\)
raises the rms deviation to \(1.5\)--\(2.0\%\), with maxima of \(4\)--\(8\%\). We take this window as the turbulent
attractor, and call the earlier snapshots the stochastization stage and the later ones decayed turbulence, as for the
wind tunnel (Sec.~\ref{sec:wt_regions}). Along the attractor \(Re_\lambda\) falls from \(87\) to \(44\) in the BS and
from \(88\) to \(29\) in the LKB runs. Only the attractor is fitted.

The DNS scaling function and the theoretical curve are matched by
\begin{equation}
    H_{\mathrm{DNS}}(\log\kappa)
    =
    \mathcal A\,
    H(\log\kappa+\xi_0),
    \label{eq:matching}
\end{equation}
where \(\mathcal A\) fixes the energy normalization and \(\xi_0\) converts the
empirical length unit into the theoretical variable. These two constants do not
alter the predicted curvature, local index, pole structure, or tail shape: the
prediction is a fixed spectral shape with fitted normalization and scale. They are
fitted by least squares in \(\log h\) over \(-1\le\log\kappa\le0.4\), separately in each time bin. The upper limit is
where the local index of the DNS spectrum falls below \(-7/2\), the steepest slope of
the theory (Fig.~\ref{fig:TheoryVsDNSCompare}c); this point moves to larger \(\kappa\) as the Reynolds number grows,
from \(\log\kappa=0.28\) at \(Re_\lambda=29\) to \(0.36\) at \(Re_\lambda=84\) and \(0.44\) after the extrapolation to
infinite Reynolds number below. Beyond it lies the dissipation range
of a simulation at finite Reynolds number, which the theory, a limit of vanishing
molecular viscosity (Sec.~\ref{sec:mellinspectrum}), does not contain. Below the lower
limit the spectrum depends steeply on the Reynolds number, and its extrapolation is not determined.

\paragraph*{Dependence on the Reynolds number.}
The collapsed spectra drift slowly during the decay (Fig.~\ref{fig:TheoryVsDNSCompare}f), and the drift is set by the
instantaneous \(Re_\lambda\). The BS and LKB runs reach a given \(Re_\lambda\) at different times and with different
ratios of the integral scale to the box. Compared at equal \(Re_\lambda\), their spectra differ in \(\log h\) by
\(0.005\) rms over the fitted range, \(0.012\) over \(0.4\le\log\kappa\le1.5\) and \(0.029\) over
\(-2.5\le\log\kappa\le-1\); compared at equal times, by \(0.028\), \(0.11\) and \(0.18\). The two initial spectra thus
lead to the same spectrum at the same Reynolds number, and the finite box plays no visible role. Along the attractor the
rms deviation from the theory over the fitted range decreases as \(Re_\lambda\) grows, from \(7.6\%\) at
\(Re_\lambda=29\) through \(5.2\%\) at \(45\) and \(3.9\%\) at \(62\) to \(2.9\%\) at \(84\), the same for the two
families (Fig.~\ref{fig:TheoryVsDNSCompare}e), while \(\xi_0\) moves from \(4.41\) to \(4.50\). The run-to-run standard
errors are below \(0.4\%\), so these deviations are systematic.

\paragraph*{Extrapolation to infinite Reynolds number.}
As for the wind tunnel (Sec.~\ref{sec:windtunnel}), we extrapolate the data, not the theory. At fixed \(\log\kappa\) we
fit \(\log h=\log h_\infty+a(\kappa)/Re_\lambda\) over the \(24\) time bins of the attractor, each at its own
\(Re_\lambda\) (Fig.~\ref{fig:TheoryVsDNSCompare}f). The extrapolation is determined for
\(-1.5\lesssim\log\kappa\lesssim1.3\): there a quadratic term in \(1/Re_\lambda\), the BS or the LKB runs alone, or the
shorter window \(10\le t\le500\) change \(\log h_\infty\) by less than \(0.1\), and by less than \(0.05\) for
\(|\log\kappa|\le1\). A fit to the \(4682\) individual snapshots of the attractor, each at its own \(Re_\lambda(t_n)\),
gives the same \(\log h_\infty\) to \(0.006\). With \(\xi_0=4.550\), the extrapolated spectrum agrees with the theory to
\(1.0\%\) rms (maximum \(3\%\)) over the fitted range, and the variants give \(1.0\)--\(1.5\%\)
(Fig.~\ref{fig:TheoryVsDNSCompare}b,e). The earliest snapshots, in the stochastization stage at
\(Re_\lambda=120\)--\(170\), are not fitted; they agree with the theory as closely, to \(1.0\)--\(1.5\%\), with
\(\xi_0=4.55\)--\(4.56\). Beyond \(\log\kappa=0.4\) the DNS falls below the theory, by \(55\%\) at \(\log\kappa=1\) for
\(Re_\lambda=84\) and by \(41\%\) after the extrapolation, which, being linear in \(1/Re_\lambda\), cannot follow a
dissipation range that recedes as the Reynolds number grows.

The same data can also be compared through the complex Mellin amplitude itself.
Being an integral over all wave numbers, it is more sensitive than the scaling
function and the structure-function index, which can be compared range by range, to
the finite size of the DNS: to the infrared shells that still carry the initial data
and to the largest resolved wave number. We postpone this comparison until the
forthcoming \(8192^3\) simulations.

\subsection{Structure function and decay laws in the DNS}
\label{sec:dns_decay}

The same simulations test two predictions that do not use the energy spectrum directly: the second-order velocity
structure function, and the decay of the energy and the enstrophy. Both were compared with the theory in
Ref.~\cite{ReviewPaperAM}; we redo the comparison here with the final data and the structure function of the odd
ensemble, because they bear directly on the classical decay laws.

\begin{figure}[htbp]
    \centering
    \includegraphics[width=\linewidth]{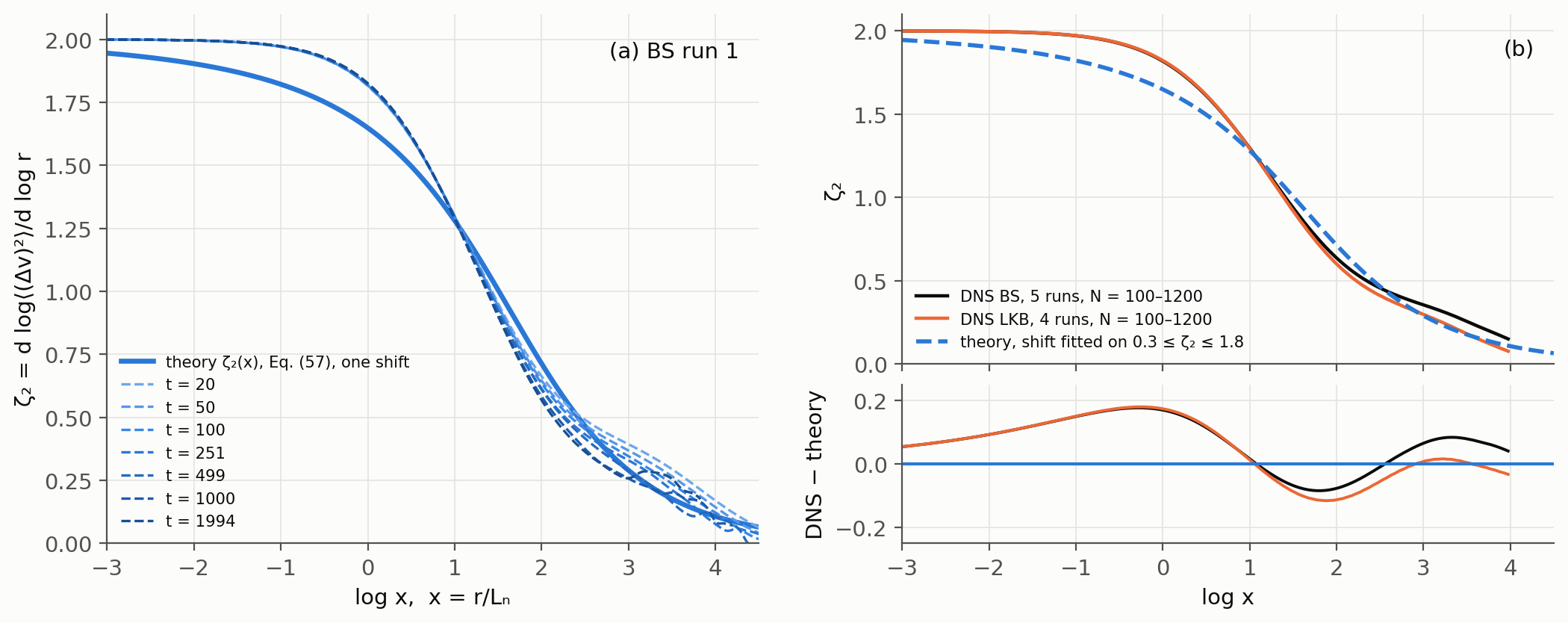}
    \caption{Effective index of the second-order structure function,
    \(\zeta_2(x)=d\log\VEV{(\Delta\vect v)^2}/d\log r\), in the \(4096^3\) DNS of Rodhiya and
    Sreenivasan~\cite{SreeniAkash2025} at \(Re_\lambda=145\), against the odd Euler ensemble~\eqref{eq:Dx}. (a) Single
    snapshots of BS run~1 from \(t\approx20\) to \(t\approx2000\), as functions of \(x=r/L_n\), and the theory with one
    shift. (b) Window means (\(N=100\)--\(1200\)) of the five BS and four LKB runs, and the theory with the shift fitted
    on \(0.3\le\zeta_2\le1.8\); bottom: DNS minus theory.}
    \label{fig:DNSindex}
\end{figure}

\paragraph*{Test of the index.}
Figure~\ref{fig:DNSindex} compares \(\zeta_2(x)\) with the DNS for the Birkhoff--Saffman (BS, \(E\propto k^2\) at small
\(k\)) and the Loitsyansky--Kolmogorov--Batchelor (LKB, \(E\propto k^4\)) initial spectra at
\(Re_\lambda=145\)~\cite{SreeniAkash2025}. The DNS index is computed from the spectra,
\(D(r)=4\sum_kE(k)\,(1-\sin kr/(kr))\), as a function of \(x=r/L_n\). The snapshots from \(t\approx20\) to \(t\approx2000\)
collapse for \(\log x\lesssim2.5\), and the window means of the two initial spectra agree within \(0.07\): the bulk of the
flow has forgotten the infrared form imposed at \(t=0\). With one fitted shift, the theory follows the DNS crossover
with rms deviation \(0.08\) (maximum \(0.17\)) for \(0.3\le\zeta_2\le1.8\). The run-to-run errors are \(0.001\), so the
deviation is systematic: the DNS crossover is narrower. At small \(x\) the DNS index approaches \(2\) as \(2-O(x^2)\), as for
any velocity field that is smooth at the dissipation scale, whereas the theory approaches \(2\) as \(2-O(x^{1/2})\): the
pole \(q=5/2\) of \(Z_\eta(q)\), the image of the \(\kappa^{-7/2}\) tail of the spectrum, adds a term \(x^{5/2}\) to \(D(x)\).
As for the spectrum, the finite-Reynolds dissipation range of the simulation (Fig.~\ref{fig:TheoryVsDNSCompare}c) is not
part of the theory. The large-separation part of the index is tested at much higher Reynolds numbers in
Sec.~\ref{sec:windtunnel}.

\paragraph*{Decay laws.}
The predicted decay laws are summarized in Sec.~\ref{sec:decaylaws}. The balance holds in the
simulations: in each of the nine runs \(-E'(t)/(2En)\) is constant to \(1\%\) over \(5<t<2000\). Table~\ref{tab:decay}
gives the exponents, fitted with a free virtual origin over \(100<t<1000\), and the moment relations with the bulk
length~\eqref{DNSLengthScale}. The BS runs decay with \((n_E,n_{En})=(1.24,2.26)\), close to the asymptotic
Euler-ensemble pair \((5/4,9/4)\); the Saffman pair \((6/5,11/5)\) lies two standard errors away. The LKB runs decay with
the Loitsyansky pair \((10/7,17/7)\); with the exponent fixed at \(9/4\) the rms residual of \(\log En\) is five to ten
times larger.

\begin{figure}[htbp]
    \centering
    \includegraphics[width=\linewidth]{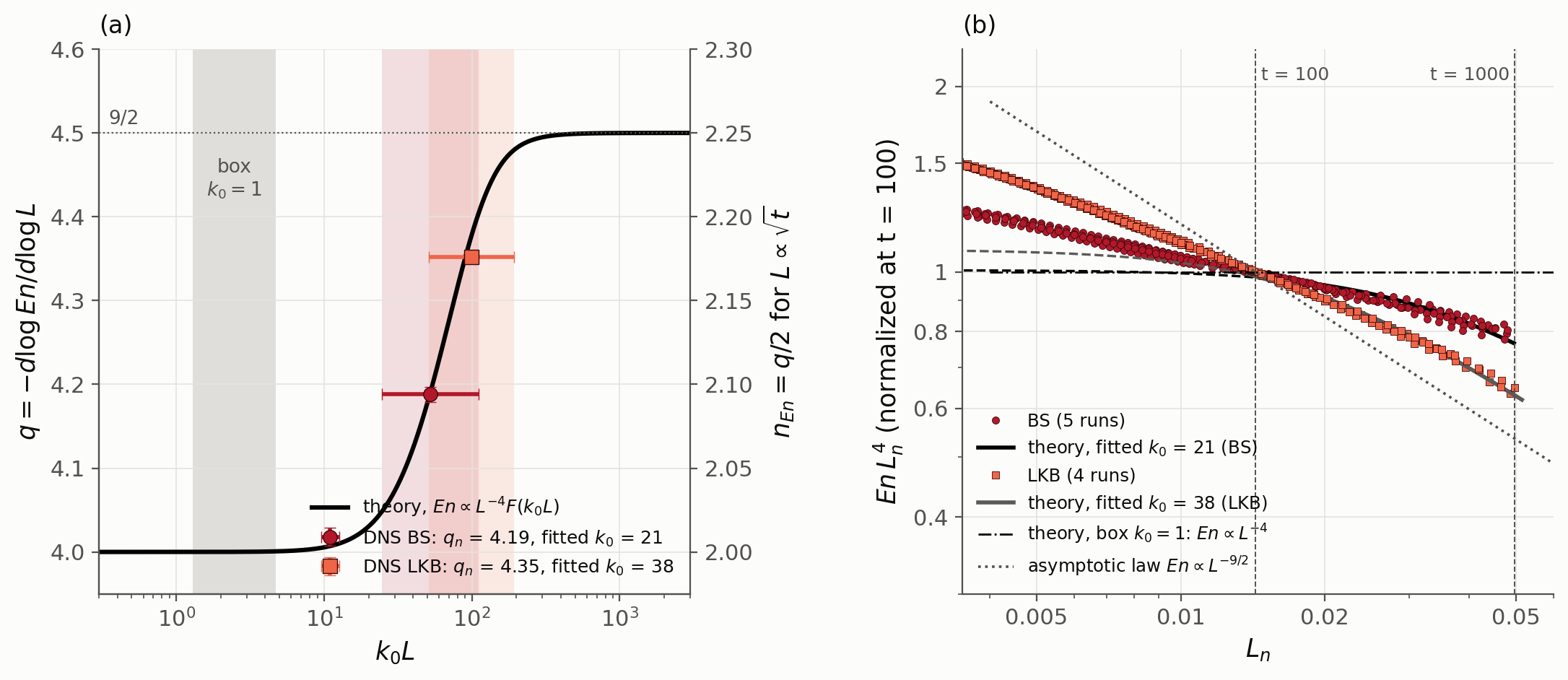}
    \caption{Decay laws in the \(4096^3\) DNS~\cite{SreeniAkash2025} against the Euler ensemble with an infrared cutoff
    \(k_0\). (a) The exponent \(q\) of \(En\propto L^{-q}\) computed from~\eqref{eq:enstrophyIntegral} as a function of
    \(k_0L\) (black; right axis: the time exponent \(n_{En}=q/2\) for \(L\propto\sqrt t\)). Shaded: the range of \(k_0L\)
    over \(100<t<1000\) for the lowest shell of the box, \(k_0=1\) (grey), and for the sharp cutoffs fitted to the BS
    (\(k_0=21\), red) and LKB (\(k_0=38\), orange) runs, with \(L=e^{\xi_0}L_n\); points: the measured \(q\) of
    Table~\ref{tab:decay}. (b) Enstrophy compensated by \(L_n^4\), normalized at \(t=100\), against the bulk
    length~\eqref{DNSLengthScale} for the nine runs, \(4.6\le t\le1000\): the laws \(A\,L^{-4}F(k_0L)\) with the fitted
    \(k_0\) for run~1 of each family (solid in the fitted range \(100<t<1000\), dashed outside), the law for the box,
    \(En\propto L^{-4}\) (dash-dotted), and the asymptotic law \(En\propto L^{-9/2}\) (dotted).}
    \label{fig:DNSdecay}
\end{figure}

The exponents of the theory depend on \(k_0L\) (Sec.~\ref{sec:decaylaws}). In the length unit of the spectral fit,
\(L=e^{\xi_0}L_n\) with \(\xi_0=4.55\), the lowest shell of the box, \(k_0=1\), gives \(k_0L=1.3\)--\(4.9\) over
\(100<t<1000\). This cutoff lies two decades below the bulk of the spectrum: the integral
in~\eqref{eq:enstrophyIntegral} equals its infinite-flow value to \(3\times10^{-4}\), and it would give
\(En\propto L^{-4}\propto(t-t_v)^{-2}\). The simulations decay faster; with the exponent fixed at \(2\) the rms residual
of \(\log En\) in the BS runs is \(4\) to \(17\) times larger than with \(9/4\). The infrared cutoff of the scaling range
is therefore not the box. We estimate it with a sharp cutoff, fitted together with the normalization,
\(En=A\,L^{-4}F(k_0L)\), where \(F\) is the
integral in~\eqref{eq:enstrophyIntegral} and \(L=e^{\xi_0}L_n\) is measured (Fig.~\ref{fig:DNSdecay}). Over \(100<t<1000\)
this gives \(k_0=21\pm1\) for the BS and \(k_0=38\pm1\) for the LKB runs, in units of the lowest shell, so that \(k_0L\)
runs over \(24\)--\(112\) and \(51\)--\(194\), the crossover of Fig.~\ref{fig:DNSdecay}a; the rms residual of \(\log En\)
is \(0.9\)--\(1.6\%\) and \(0.7\)--\(1.5\%\), against \(0.3\)--\(1.3\%\) for a power law in \(L_n\). The fitted cutoff
lies at \(\log(k_0L_n)=-1.4\)--\(0.2\) (BS) and \(-0.6\)--\(0.7\) (LKB): at the lower edge of the fitted bulk
(Sec.~\ref{sec:dns_collapse}), where the spectrum leaves the theory and still carries the initial data. This infrared
part of the spectrum carries \(50\)--\(70\%\) of the energy but only \(6\)--\(8\%\) of the enstrophy, and the energy
balance ties the decay of the bulk to that of the energy. The fit has limits. The fitted law steepens across the window,
from \(q=4.06\) to \(4.36\) (BS) and from \(4.19\) to \(4.48\) (LKB), whereas the measured enstrophy keeps
\(En\propto L_n^{-4.19}\) and \(L_n^{-4.35}\) with no visible curvature; over the whole attractor,
\(4.6\le t\le817\), the fitted \(k_0\) doubles and the rms residual rises to \(5\)--\(6\%\). In time, the measured
exponents exceed \(q/2\) because the bulk length grows faster than diffusively, \(L_n\propto t^{0.54}\) (BS) and
\(t^{0.56}\) (LKB): with the measured \(L_n\) the fitted law reproduces \(En(t)\), but with \(L\propto\sqrt t\) no cutoff
gives an exponent above \(9/4\). Fitted in time alone, with \(k_0^2\tilde\nu\) free, the law reproduces the BS decay on
the attractor as well as a free power law, but only with \(k_0L\approx60\)--\(950\), which the measured position of the
bulk excludes, and it cannot reproduce the LKB exponent \(2.40\). The LKB decay thus needs, besides the cutoff, the faster
growth of the bulk length, an infrared effect of the \(k^4\) initial data rather than of the box. These fits use the sharp
cutoff of~\eqref{eq:enstrophyIntegral}, which is only an estimate: a finite flow cuts the spectrum off with a smooth
function, set by its boundaries and its initial data, and the crossover of an integral dominated by its lower limit
depends on that function. The fits locate the effective cutoff; the test of the theory is the bulk of the spectrum
(Sec.~\ref{sec:dns_collapse}).

Ishida, Davidson and Kaneda found that for \(k^4\) initial spectra in domains much larger than the integral scale the
Loitsyansky integral is approximately conserved~\cite{Ishida2006}, and both self-similar and non-self-similar decay have
been found in simulations with other initial conditions~\cite{Matsumoto2025}. At \(Re_\lambda=145\) the infrared initial
data thus set the decay rates, while the shape of the bulk, which carries the structure-function index, is the same for
both initial spectra (Fig.~\ref{fig:DNSindex}).

\begin{table}[htbp]
\centering
\caption{Decay exponents of the \(4096^3\) DNS~\cite{SreeniAkash2025}: fits \(E,En\propto(t-t_v)^{-n}\) with a free virtual
origin \(t_v\) over \(100<t<1000\), and the moment relations \(En\propto L_n^{-q}\), \(L_n\propto t^{a}\) over the same range,
with the bulk length~\eqref{DNSLengthScale}; with \(L_M=\int kE\,dk/\int k^2E\,dk\) instead, \(q=4.29\pm0.02\) and
\(4.52\pm0.01\), \(a=0.526\) and \(0.535\). Errors are standard errors over the runs. The exponents of the Euler ensemble
depend on the infrared cutoff \(k_0L\) (Sec.~\ref{sec:decaylaws}): \(k_0L=1.3\)--\(4.9\) for the lowest shell of the box,
\(24\)--\(112\) (BS) and \(51\)--\(194\) (LKB) for the fitted cutoffs.}
\label{tab:decay}
\begin{tabular}{lcccc}
\hline
Initial spectrum (runs) & \(n_E\) & \(n_{En}\) & \(q\) & \(a\) \\
\hline
BS, \(E\propto k^2\) (5) & \(1.242\pm0.020\) & \(2.256\pm0.020\) & \(4.19\pm0.01\) & \(0.539\pm0.001\) \\
LKB, \(E\propto k^4\) (4) & \(1.417\pm0.007\) & \(2.401\pm0.006\) & \(4.35\pm0.01\) & \(0.555\pm0.001\) \\
\hline
Euler ensemble, \(k_0L\lesssim10\) & \(1\) & \(2\) & \(4\) & \(1/2\) \\
Euler ensemble, \(k_0L\gtrsim400\) & \(5/4\) & \(9/4\) & \(9/2\) & \(1/2\) \\
Saffman & \(6/5\) & \(11/5\) & & \\
Loitsyansky & \(10/7\) & \(17/7\) & & \\
\hline
\end{tabular}
\end{table}

\paragraph*{Initial data.}
The simulations start from random Fourier modes with a prescribed spectral envelope, which contains a \(k^{-5/3}\)
range~\cite{SreeniAkash2025}. This range degrades quickly. Such initial data are appropriate for a test of an
attractor, which must not depend on them: the comparison uses only the snapshots in which the scaling collapse holds,
the window \([N_0,N_1]\) of Sec.~\ref{sec:dns_match}, and its stability under a change of the window is the check. The
attractor itself has no extended \(k^{-5/3}\) range but a continuously varying local index~\cite{NTDT2}, so the
initial \(k^{-5/3}\) range is not expected to survive. Earlier simulations of freely decaying turbulence did not
develop a clear \(k^{-5/3}\) range either~\cite{Ishida2006}, and in the freely decaying turbulent blob of Matsuzawa
\emph{et al.} the \(\kappa^{-5/3}\) scaling disappears once the initial period has passed~\cite{Matsuzawa2026}.

\section{Comparison with the Max Planck wind tunnel}
\label{sec:windtunnel}

The DNS reach \(Re_\lambda\approx145\). Much higher Reynolds numbers are available in grid turbulence. K\"uchler,
Bewley and Bodenschatz measured the second-order structure function \(S_2(r)\) of decaying turbulence in the Max Planck
Variable Density Turbulence Tunnel, for \(11\) runs with \(Re_\lambda=413\)--\(5779\), at one station \(8.3\)~m
downstream of an active grid~\cite{GregXi2}. Here we summarize the comparison of these data with the structure
function~\eqref{eq:Dx} of the odd Euler ensemble. The details are given in Ref.~\cite{MigdalWindTunnel2026}, and the
code and data tables in Refs.~\cite{ZenodoNTDT3,LoopEquationsRepo}.

\subsection{Measured index and the extrapolation to infinite Reynolds number}

The measured index is \(\alpha_{\rm exp}(r)=d\log S_2/d\log r\), so the unknown normalization of \(S_2\) drops out. The
theory predicts it as the longitudinal index of~\eqref{eq:DLL}, a function of \(x=r/L(t)\) alone. Each run is a separate flow measured at one
decay time, so it supplies a single value of \(L\). We relate it to the measured integral scale
\(\ell=\int_0^{r_0}R\,dr\), \(R=1-S_2/2u'^2\), which is \(0.1\)--\(0.4\)~m in these runs, by the ratio
\(s=\log(L/\ell)\). At the smallest separations the measured index reaches the dissipative limit \(\alpha\to2\), the
\(r^2\) behaviour of a smooth velocity field, which the theory contains as the leading pole \(q=2\)
of~\eqref{eq:Dx}: \(\alpha_{\rm exp}=2.00\) at \(r/\eta\approx0.3\)--\(0.4\) in the runs with \(Re_\lambda\le1046\), and
\(1.94\)--\(1.99\) at the smallest measured \(r/\eta=0.5\)--\(2\) of the higher-Reynolds runs.

\begin{figure}[htbp]
    \centering
    \includegraphics[width=\linewidth]{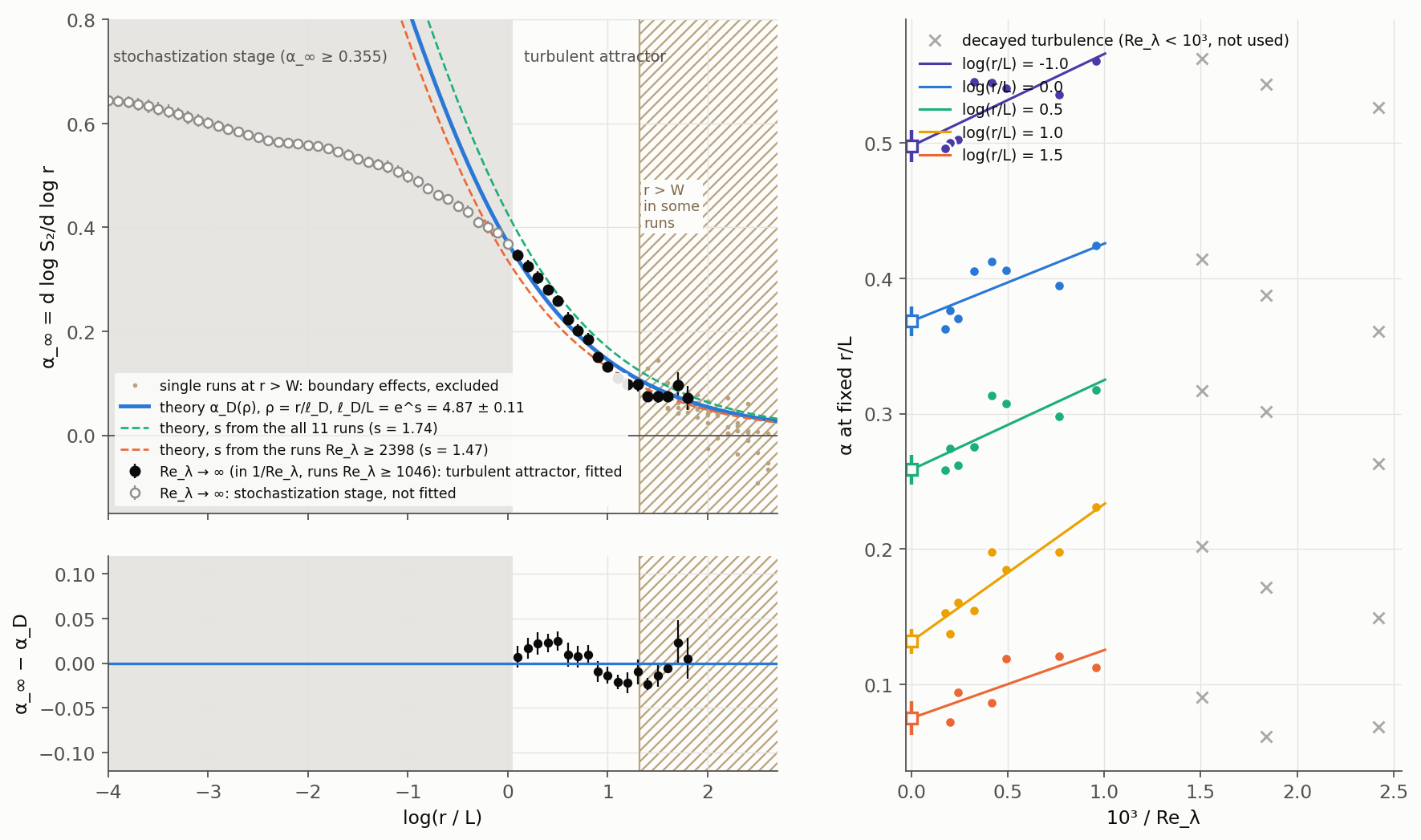}
    \caption{Max Planck wind-tunnel data~\cite{GregXi2} against the odd Euler ensemble~\cite{MigdalWindTunnel2026}.
    Left: the index extrapolated to \(Re_\lambda\to\infty\) in \(1/Re_\lambda\) (runs \(Re_\lambda\ge1046\)) as a
    function of \(\log(r/\ell)\), against the longitudinal index of the theory, Eq.~\eqref{eq:DLL}, with one scale
    parameter (blue). In the figure
    \(L\) denotes the integral scale \(\ell\), and \(\ell_D\) the diffusion length \(L(t)\). Filled circles: the
    turbulent attractor, to which the theory is fitted; open grey circles: the stochastization stage (grey band). The
    hatched band starts where the first run reaches \(r=W\); the small dots there are the excluded single-run values.
    Dashed lines use the shifts fitted to the extrapolations with all runs and with \(Re_\lambda\ge2398\). Bottom:
    residuals. Right: \(\alpha_{\rm exp}\) against \(1/Re_\lambda\) at fixed \(r/\ell\); squares are the extrapolated
    values, grey crosses the decayed-turbulence runs, which are not used.}
    \label{fig:WTextrap}
\end{figure}

The theory describes the limit \(Re_\lambda\to\infty\), so we extrapolate the data, not the theory. At fixed
\(\log(r/\ell)\) we fit \(\alpha_{\rm exp}=\alpha_\infty+c/Re_\lambda\) over the eight runs with \(Re_\lambda\ge1046\). The
longitudinal index of the theory, with \(\log x=\log(r/\ell)-s\), is then fitted to the extrapolated index in the
turbulent attractor defined below, with weights from the extrapolation. A single parameter, \(s=1.583\pm0.023\)
(\(L/\ell=4.87\pm0.11\)), gives \(\chi^2/\mathrm{dof}=2.9\) and a weighted rms deviation of \(0.013\), with residuals within
\(\pm0.03\) (Fig.~\ref{fig:WTextrap}). As the lower-Reynolds runs are dropped, \(s\) decreases to \(1.47\) and
\(\chi^2/\mathrm{dof}\) to \(1.3\); with all eleven runs \(s=1.74\). Together with the cut of the tail
(\(s=1.46\)--\(1.58\)), an extrapolation in \(Re_\lambda^{-1/2}\) instead of \(1/Re_\lambda\) (\(s=1.41\),
\(\chi^2/\mathrm{dof}=3.2\)) and a bootstrap over the runs (\(s=1.56\pm0.04\)), this gives a systematic uncertainty of
about \(\pm0.1\) on \(s\), larger than the statistical one. The even ensemble gives an index within \(0.004\) of the odd
one in this range, so the data do not resolve the parity.

The shape of the large-separation index is specific to the theory. Von K\'arm\'an spectra
\(E\propto k^m(1+k^2)^{-(m+5/3)/2}\), fitted with the same single scale parameter to the same extrapolated index and with
the same weights, give \(\chi^2/\mathrm{dof}=13\) for a Saffman (\(m=2\)) and \(24\) for a Batchelor (\(m=4\)) infrared
range, against \(2.9\). The measured index falls as \(r^{-1.03}\). The theory gives a \(1/r\) tail, because its scaling
function \(H(\kappa)\) is finite at \(\kappa=0\); spectra of the K41 type with a \(k^2\) or \(k^4\) infrared range give
steeper tails.

\subsection{Turbulent attractor, stochastization stage, decayed turbulence and boundary effects}
\label{sec:wt_regions}

\begin{figure}[htbp]
    \centering
    \includegraphics[width=\linewidth]{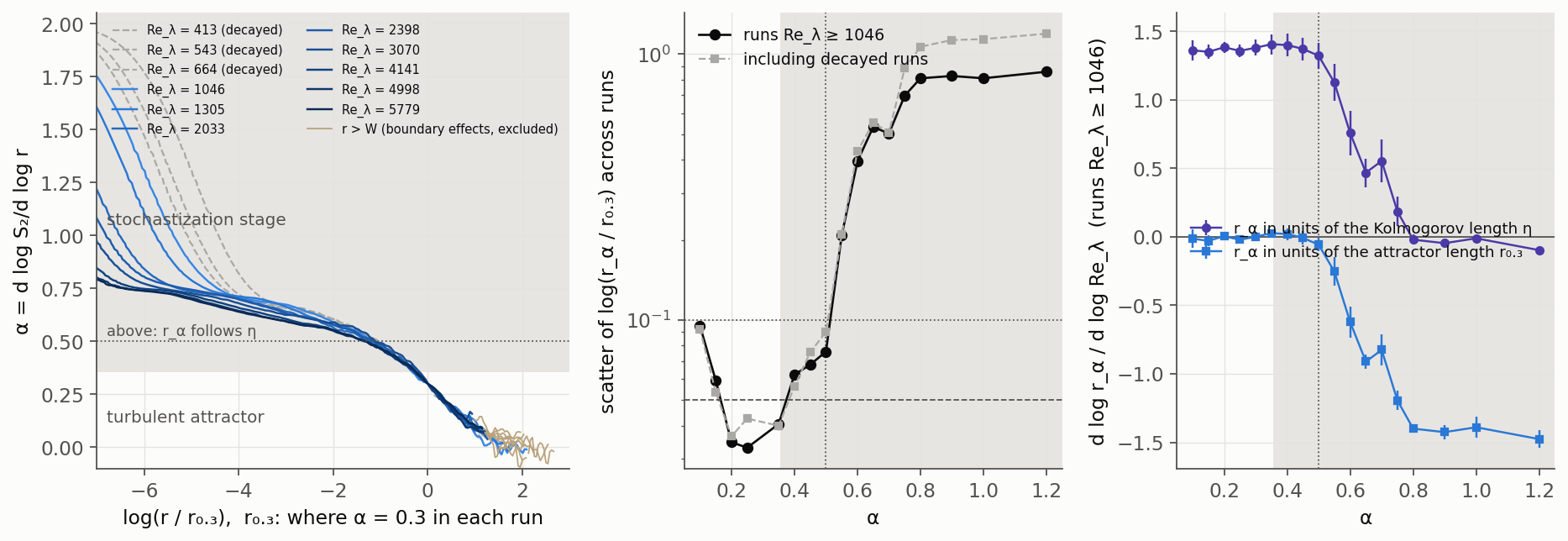}
    \caption{Turbulent attractor and stochastization stage in the Max Planck
    data~\cite{GregXi2,MigdalWindTunnel2026}. Left: the measured index of each run against \(\log(r/r_{0.3})\), where
    \(r_{0.3}\) is the separation at which that run's index equals \(0.3\); dashed grey curves are the decayed-turbulence
    runs, \(Re_\lambda<10^3\), and the tan ends are the boundary-effects region \(r>W\). Middle: the scatter of
    \(\log(r_\alpha/r_{0.3})\) across the runs at fixed \(\alpha\). Right: the Reynolds-number dependence of
    \(r_\alpha\) in units of the Kolmogorov length and of \(r_{0.3}\) (runs \(Re_\lambda\ge1046\)). In the turbulent
    attractor, \(\alpha<0.355\) (white), the runs collapse tightly; in the stochastization stage (grey) the scatter
    grows, slowly up to \(\alpha\approx0.5\) (dotted lines) and steeply above, where \(r_\alpha\) follows \(\eta\).}
    \label{fig:WTattractor}
\end{figure}

The theory applies after the initial data have been forgotten, where the index depends on \(r\) only through one
length that grows as \(\sqrt t\). Since all runs are measured at one station, this is tested across the runs at fixed
index. Let \(r_\alpha\) be the separation at which a run's index crosses the level \(\alpha\). Where the attractor
holds, \(\log(r_\alpha/r_{0.3})\) is the same for all runs (Fig.~\ref{fig:WTattractor}). This test separates four
regions of the data.
\begin{enumerate}
    \item The \emph{turbulent attractor}, \(\alpha<0.355\), roughly \(\ell\lesssim r\le W\). The scatter of
    \(\log(r_\alpha/r_{0.3})\) across the runs with \(Re_\lambda\ge1046\) stays below \(0.05\), with no drift with the
    Reynolds number. The theory is fitted here.
    \item The \emph{stochastization stage}, \(\alpha\ge0.355\). The scatter grows slowly up to \(\alpha=0.5\), to
    \(0.08\), and then jumps, to \(0.21\) at \(\alpha=0.55\) and above \(0.8\) for \(\alpha\ge0.8\). There \(r_\alpha\)
    follows the Kolmogorov length, \(|d\log(r_\alpha/\eta)/d\log Re_\lambda|\le0.1\), as in the Reynolds-independent
    inertial range found by K\"uchler \emph{et al.}~\cite{GregXi2}: the flow still carries the inertial range produced
    by the grid.
    \item \emph{Decayed turbulence}: the three runs with \(Re_\lambda<10^3\), too weak for the asymptotic theory. They
    are not used in the extrapolation.
    \item The \emph{boundary-effects region}, \(r>W\). A separation longer than the width of the flow cannot probe
    isotropic turbulence. The width \(W=1.58\)~m follows from the finite-size oscillations of \(\alpha_{\rm exp}\)
    at the largest separations, which the longitudinal index of the theory reproduces by removing the wave numbers
    \(\kappa<\pi L/W\)~\cite{MigdalWindTunnel2026}; it agrees with the \(1.5\)~m width of the test sections of the
    tunnel~\cite{Bodenschatz2014}. These points, \(16\)--\(38\%\) of each tail, are excluded. The flow as a whole never
    reaches \(\ell\gtrsim W\): \(\ell\le0.27\,W\) in every run.
\end{enumerate}

In the stochastization stage the measured index falls more slowly than the theory: at \(r=\ell/e\) it is
\(\alpha_\infty=0.50\pm0.01\) against the predicted \(0.82\). This is not a finite-size effect, since the finite-width
correction changes the index by at most \(0.02\) there against observed differences of \(0.3\)--\(1.2\). Nor is it a
finite-Reynolds correction: at fixed \(r/\ell<1\) the index moves \emph{away} from the theory as \(Re_\lambda\) grows,
because the inertial range, which scales with \(\eta\), widens as \(\ell/\eta\propto Re_\lambda^{3/2}\). We attribute
it to the finite age of the flow. All runs are measured at one station, with nearly the same decay time scale
\(\tau=u'^2/\varepsilon=0.8\)--\(1.4\)~s, and the integral scale stays approximately constant along the test
section~\cite{GregXi2}; the elapsed decay time is not varied in this data set. The self-similar diffusive state has
therefore not been reached at small \(x\), which at fixed \(r\) is reached only after the longest decay times; in the
DNS, too, the snapshots collapse only after a long initial transient (Fig.~\ref{fig:DNSindex}). This interpretation is
testable. Measured at several stations, with the attractor region defined by the same fixed-\(\alpha\) collapse before
any comparison with the theory, the region should extend to smaller \(r/\ell\) as the elapsed decay time grows, and the
level \(\alpha^\ast\) that bounds it should rise above \(0.355\).

\section{Decay laws in grid turbulence and in a turbulent blob}
\label{sec:blob}

Decay exponents near Saffman's \(6/5\) are frequently reported for grid turbulence, which has been interpreted as
Saffman turbulence~\cite{KrogstadDavidson2010,Sinhuber2015}; Saffman decay has also been generated with magnetic
stirrers~\cite{GorceFalcon2024}. The classic measurement of Comte-Bellot and Corrsin gave
\(1.25\pm0.01\)~\cite{GridTurbulence_1966}, the asymptotic Euler-ensemble value \(5/4\), which requires an infrared
cutoff at the bulk of the spectrum (Sec.~\ref{sec:decaylaws}). Fitted decay exponents depend on the
virtual origin, on the fitting range and, through the energy integral, on the infrared part of the spectrum. The test
of Sec.~\ref{sec:wt_regions} does not use them: it compares the shape of the bulk index at one station, with one
scale parameter.

The isolated turbulent blob of Matsuzawa \emph{et al.}~\cite{Matsuzawa2026} removes the walls and the mean flow at
the start of the decay. Homogeneity and local isotropy, however, are strongly violated by the radial dependence of the flow. At \(Re_\lambda=60\)--\(200\) its energy first decays with the exponent \(1.3\), with a fitted
virtual origin, above Saffman's \(6/5\) and the asymptotic Euler-ensemble value \(5/4\). Once the integral scale saturates
at about a quarter of the tank, the energy decays as \(t^{-2}\), until \(Re_\lambda\) falls to about \(10\). The
authors note that the integral scale grows only as \((t-t_0)^{0.16\pm0.06}\), slower than both the \(t^{2/5}\) of
Saffman's theory and the \(t^{1/2}\) they attribute to the loop-equation solution~\cite{migdal2024quantum}. In the
present solution, however, the length that grows as \(\sqrt{t+t_0}\) is the diffusion length \(L(t)\), which sets the bulk
of the spectrum, not the integral scale, which weights the smallest wave numbers: in the blob it starts at the blob
radius and saturates at the tank scale~\cite{Matsuzawa2026}.

\begin{figure}[tbp]
    \centering
    \begin{minipage}[t]{\linewidth}
        \centering
        \includegraphics[width=0.77\linewidth]{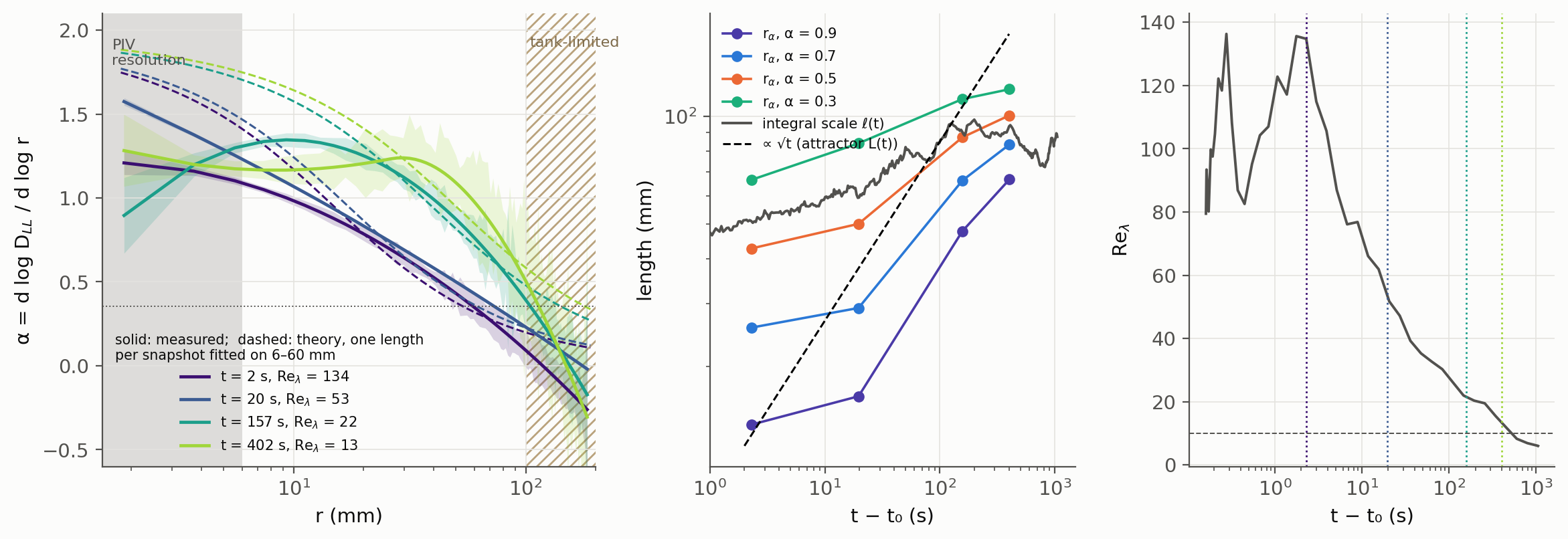}
        \par\vspace{1mm}{\small\textbf{(a)}}
    \end{minipage}
    \begin{minipage}[t]{\linewidth}
        \centering
        \includegraphics[width=0.77\linewidth]{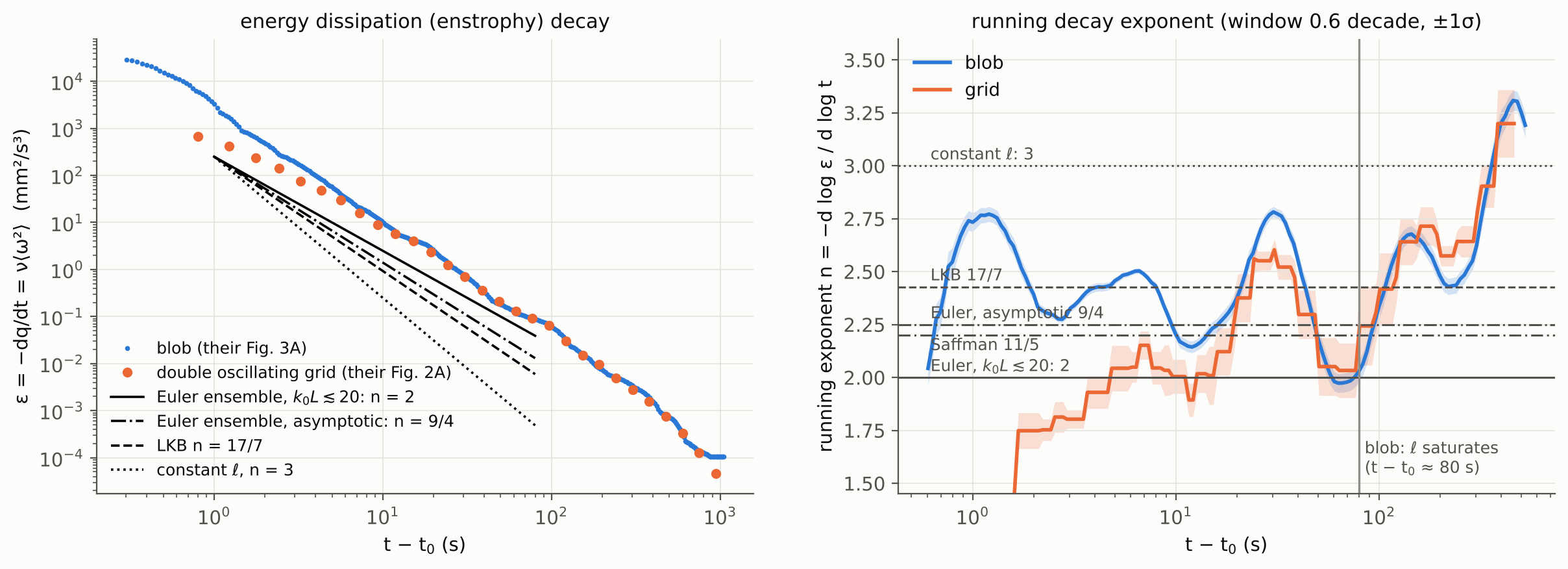}
        \par\vspace{1mm}{\small\textbf{(b)}}
    \end{minipage}
    \begin{minipage}[t]{\linewidth}
        \centering
        \includegraphics[width=0.77\linewidth]{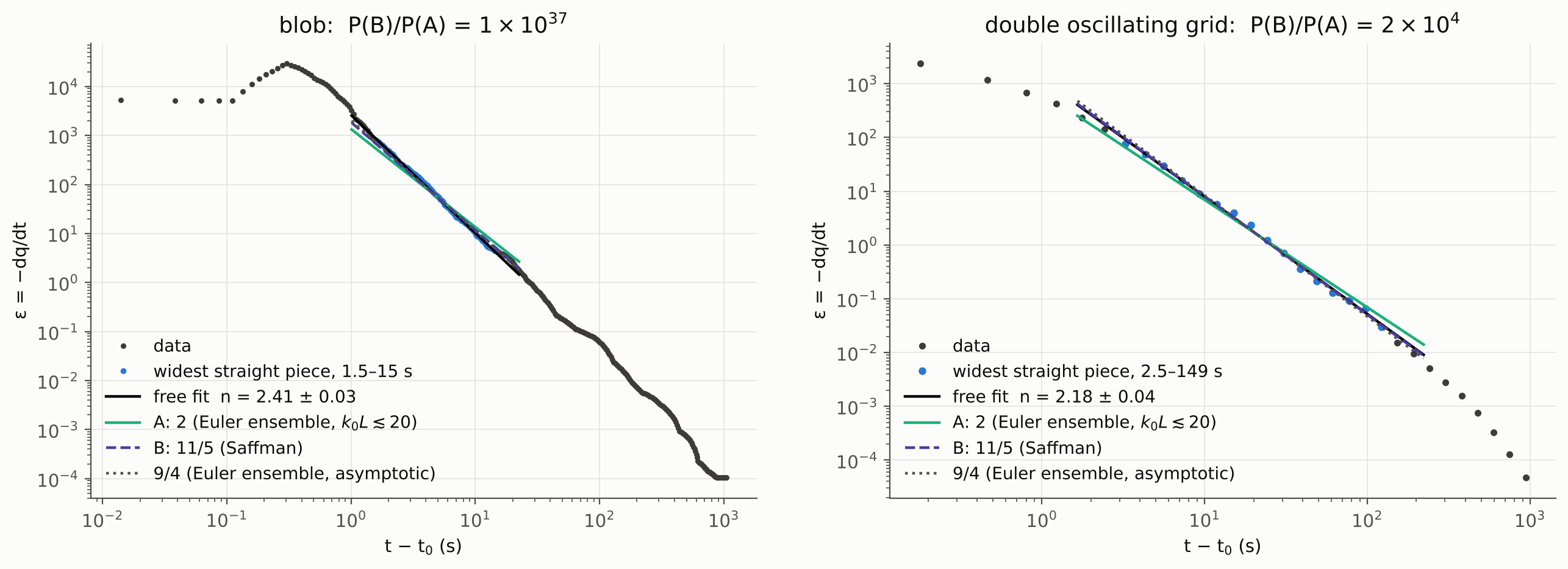}
        \par\vspace{1mm}{\small\textbf{(c)}}
    \end{minipage}
    \caption{The turbulent blob of Matsuzawa \emph{et al.}~\cite{Matsuzawa2026}, from their published
    data~\cite{MatsuzawaData2026}. (a) Left: the local index \(\alpha=d\log D_{LL}/d\log r\) of the measured structure
    function at four times (solid, \(\pm1\sigma\) bands) and the best one-length fit of the longitudinal index of the
    Euler ensemble, Eq.~\eqref{eq:DLL}, on \(r=6\)--\(60\)~mm (dashed); grey: below the resolution of the particle image velocimetry; hatched: tank-limited
    separations. Centre: the lengths \(r_\alpha\) at fixed index and the integral scale \(\ell(t)\), against the growth
    \(\propto\sqrt t\) of the diffusion length \(L(t)\) of the turbulent attractor. Right: \(Re_\lambda(t)\) and the four
    snapshots. (b) Left: the dissipation rate \(\varepsilon=-dq/dt=\nu\VEV{\omega^2}\) of the blob and of their
    oscillating-grid turbulence, with the slopes \(2\) (Euler ensemble, \(k_0L\lesssim20\)), \(9/4\) (Euler ensemble,
    asymptotic), \(17/7\) (Loitsyansky--Kolmogorov--Batchelor) and \(3\) (constant integral scale). Right: the running
    decay exponent \(-d\log\varepsilon/d\log t\) in a window of \(0.6\) decade, with the Euler-ensemble (\(2\) and
    asymptotic \(9/4\)), Saffman, Loitsyansky and constant-\(\ell\) values; the vertical line marks the saturation of \(\ell\)
    at \(t-t_0\approx80\)~s. (c) The test of the hypotheses A, \(\varepsilon\propto(t-t_0)^{-2}\) (Euler ensemble,
    \(k_0L\lesssim20\)), and B, \(\varepsilon\propto(t-t_0)^{-11/5}\) (Saffman), on the widest straight piece of the
    log--log data (blue points): straight-line fit (black), the two hypotheses and the asymptotic Euler-ensemble law
    \(\varepsilon\propto(t-t_0)^{-9/4}\) (dotted), with fitted intercepts. Left: the blob, likelihood ratio
    \(\Lambda_{B/A}=1\times10^{37}\). Right: the oscillating grid, \(\Lambda_{B/A}\approx2\times10^4\). The panels write
    \(\Lambda_{B/A}\) as \(P(B)/P(A)\).}
    \label{fig:Matsuzawa}
\end{figure}

We applied the tests of this paper to their published data~\cite{MatsuzawaData2026} (Fig.~\ref{fig:Matsuzawa}): four
snapshots of the longitudinal structure function of the blob from planar particle image velocimetry, \(r=2\)--\(184\)~mm,
at \(t-t_0\approx2\), \(20\), \(157\) and \(400\)~s after the end of the forcing, while \(Re_\lambda\) falls from \(134\) to
\(13\), and the dissipation rate \(\varepsilon=-dq/dt\) at \(330\) times. The early profiles still contain an
inertial-like stretch, \(\alpha\approx0.7\)--\(1\) over a decade of \(r\): a stochastization stage. At late times the tail
\(\alpha<0.355\) lies at \(0.2\)--\(0.5\) of the tank and falls three to four times faster than the theory, a boundary
effect. The lengths \(r_\alpha\) at fixed index grow as \(t^{0.1}\)--\(t^{0.3}\), like the integral scale, not as \(\sqrt t\),
and a single length per snapshot fits the measured index, with the longitudinal index of the theory, only to an rms
deviation of \(0.09\)--\(0.23\), against \(0.013\)
in the turbulent attractor of the wind tunnel. At the smallest separations the measured index stays at \(1.3\)--\(1.6\)
even where \(r<\eta\): the measurement does not resolve the dissipative limit \(\alpha\to2\) that the wind-tunnel data
reach.

The enstrophy is less sensitive to the boundaries than the energy, since it is carried by the small scales. In
homogeneous turbulence \(\nu\VEV{\omega^2}=\varepsilon\), and the decay laws \(E\propto t^{-n_E}\) give
\(\varepsilon\propto t^{-n_E-1}\): Saffman's \(11/5\), Loitsyansky's \(17/7\), and for the Euler ensemble \(2\) at
\(k_0L\lesssim10\), rising to the asymptotic \(9/4\) at \(k_0L\gtrsim400\) (Sec.~\ref{sec:decaylaws}). The one-length fits
of the index give \(L\approx250\)--\(290\)~mm in the early snapshots (Fig.~\ref{fig:Matsuzawa}a), so that
\(k_0L\lesssim20\) for a cutoff set by any length between the blob diameter, about twice the integral scale
\(\ell\approx50\)--\(60\)~mm, and the tank: the theory predicts \(\varepsilon\propto(t-t_0)^{-2}\). Here, however, \(\varepsilon\) is the decay rate \(-d\VEV{q}/dt\) of the energy averaged over the
field of view~\cite{Matsuzawa2026}. While the blob expands, this includes the turbulent transport out of the field of
view (Eq.~(2) of Ref.~\cite{Matsuzawa2026}), so it equals \(\nu\VEV{\omega^2}\) only where that transport is negligible.
The running exponent \(-d\log\varepsilon/d\log t\) (Fig.~\ref{fig:Matsuzawa}b) is too noisy to
decide, so we compare the hypotheses directly on the log--log data. Loitsyansky's law requires an initial spectrum tuned to cancel the leading \(k^2\)
term at small \(k\). No such tuning was made in this experiment, so the only classical alternative to the Euler-ensemble
exponent \(2\) (hypothesis A) is Saffman's \(11/5\) (hypothesis B). We take the widest straight piece of
\(\log\varepsilon\) against \(\log(t-t_0)\), the widest window in which a straight line fits within twice the
point-to-point noise of the data. For the blob it is \(t-t_0=1.5\)--\(15\)~s, one decade, before the integral scale
saturates. A straight-line fit there gives \(n=2.41\pm0.03\); the error accounts for the correlation of the residuals, which
leaves about nine independent points, but not for the uncertainty of the virtual origin, which we take from
Ref.~\cite{Matsuzawa2026}. With the slope fixed and the intercept free, the likelihood ratio of the two hypotheses is
\(\Lambda_{B/A}=1\times10^{37}\) (Fig.~\ref{fig:Matsuzawa}c), and \(2\times10^{45}\) and \(5\times10^{50}\) for
straightness thresholds of \(3\) and \(1.5\) times the noise. This ranks Saffman's law far above the Euler-ensemble
prediction, but neither describes the window: the measured exponent lies \(15\sigma\) above \(2\), \(7\sigma\) above
\(11/5\) and \(5\sigma\) above the asymptotic \(9/4\). With a sharp cutoff \(k_0\) left free, as for the DNS, the law
\(A\,(t-t_0)^{-2}F(k_0L)\) cannot reach this exponent, which exceeds the maximum \(9/4\) of the theory: its best fit is the
asymptotic law, with two to three times the rms residual of a free power law. Numerically the measured exponent agrees
with Loitsyansky's \(17/7=2.43\), a law that does not apply here, as explained above. The window is
short and early, while \(Re_\lambda\) falls from \(134\) to about \(60\), and the transport out of the field of view may
contribute to the excess. For their oscillating-grid turbulence, \(31\) points, the
widest straight piece, \(2.5\)--\(150\)~s, gives \(n=2.18\pm0.04\) and \(\Lambda_{B/A}\approx2\times10^4\), and
\(4\times10^4\) for the threshold \(3\), but \(3\times10^{-9}\) for the threshold \(1.5\), whose window, \(0.8\)--\(30\)~s,
gives \(n=1.95\pm0.04\): these data do not discriminate. With a free sharp cutoff, the law \(A\,(t-t_0)^{-2}F(k_0L)\) fits the
widest straight piece slightly better than a power law, with rms residual \(0.12\) against \(0.13\), as \(k_0L\) rises
from \(40\) to \(250\) across the window. The
enstrophy integral \(\nu\int k^2E\,dk\) of the measured spectra is dominated by the high-wave-number noise of the
measurement after the first few seconds, so the spectra themselves cannot be used for this test.

The blob is too small, and too inhomogeneous in the radial direction, for the attractor
of homogeneous turbulence in infinite space, and its Reynolds numbers are small; its structure functions cannot resolve
a turbulent attractor in the effective index. Only the global time dependence can be compared, and only roughly: the
dissipation decays faster than both the Euler-ensemble prediction for this geometry, \(\varepsilon\propto t^{-2}\), and
Saffman's law, and faster than the maximum \(9/4\) that the theory reaches with any infrared cutoff. A test
of the index itself needs a much larger experiment, with the parameters of the Max Planck facility: a large tunnel,
Reynolds numbers about \(40\) times larger, \(Re_\lambda\sim5000\) against \(134\), and a longer stochastization time
before the measurement, so that the turbulent attractor has time to form.

\section{Discussion}
\label{sec:discussion}

The comparisons with data give one picture. The turbulent attractor is a property of the infinite system at infinite
Reynolds number, and a flow approaches it only where its boundaries, its discretization, its finite Reynolds number and
its initial data cease to matter. In the final \(4096^3\) DNS the collapsed spectrum depends only on the instantaneous
Reynolds number, the same for the Saffman and Loitsyansky initial spectra, and its deviation from the theory decreases as
the Reynolds number grows; extrapolated to infinite Reynolds number it agrees with the theory to \(1\%\) in the bulk of the
enstrophy range. The structure-function index deviates systematically, by up to \(0.17\), through the finite-Reynolds
dissipation range, and the decay rate depends on the infrared initial data: exponents close to the asymptotic
Euler-ensemble pair \((5/4,9/4)\) for the Saffman runs and Loitsyansky's for the \(k^4\) runs, with the energy balance
exact. Only the bulk of the spectrum is a reliable prediction of the scaling function: the energy integral diverges in
the infrared, and the global integrals, energy and enstrophy, depend on how a finite flow cuts off the spectrum, which a
sharp cutoff only estimates. Estimated in this way, the cutoff lies at the lower edge of the bulk, where the spectrum
still carries the initial data, not at the lowest shell of the box, and the asymptotic decay laws would set in only when
it reaches the bulk. The Max Planck wind tunnel at \(Re_\lambda\) up
to \(5779\) displays a turbulent-attractor region in which the longitudinal index extrapolated to infinite Reynolds number
matches the odd ensemble with one scale parameter, while von K\'arm\'an forms do not; the departures at smaller
separations belong to an earlier stochastization stage that still carries the inertial range of the grid. The turbulent
blob of Matsuzawa \emph{et al.}~\cite{Matsuzawa2026} is too small, too inhomogeneous and too weak to display the
attractor. A decisive test of the index needs a larger experiment with the parameters of the Max Planck facility and a
longer stochastization time, and the forthcoming \(8192^3\) simulations will allow a comparison through the complex
Mellin amplitude of the spectrum itself. The log-periodic oscillations generated by the Riemann zeros~\cite{NTDT2} remain a
theoretical prediction: their contribution to the effective index is far too small for present experiments and
simulations~\cite{MigdalWindTunnel2026}.

\section*{Conflict of Interest Statement}
The authors declare  that the research was conducted in the absence of any commercial or financial relationships that could be construed as a potential conflict of interest.

\section*{Author Contributions}
% TODO (authors): describe the contributions of each author.
AM and WW contributed to this work. Both authors have approved it for publication.

\section*{Funding}
AM research was self-funded. 
WW  was supported partly by the NSF award 24-600, Number 2534977.

\section*{Generative AI Statement}
AM used generative AI in preparing this manuscript, for the numerical analyses as well as for the text. The
theory and its derivations are the author's own work, as are the Mathematica notebooks of
Refs.~\cite{DecayTurb23,MB40,MB41,MB42,MB43,MB44,MB45,MB46,MBMellinDT} and of Ref.~\cite{LoopEquationsRepo}. The Python code
was written and run by Claude Opus 5.5 (Anthropic), working as a coding assistant under the WW direction: the code
of Refs.~\cite{ZenodoNTDT3,LoopEquationsRepo} for the wind-tunnel comparison of Ref.~\cite{MigdalWindTunnel2026}, and the
code for:
\begin{itemize}
    \item the analysis of the \(4096^3\) DNS spectra: the selection of the turbulent attractor, the scaling-function
    collapse and its extrapolation to infinite Reynolds number, the structure-function index, the energy balance and the
    decay exponents and the fit of the infrared cutoff (Figs.~\ref{fig:TheoryVsDNSCompare}, \ref{fig:DNSindex}
    and~\ref{fig:DNSdecay}, Table~\ref{tab:decay});
    \item the reanalysis of the Max Planck wind-tunnel data with the longitudinal structure function, and the
    comparison with alternative crossover forms (Figs.~\ref{fig:WTextrap} and~\ref{fig:WTattractor});
    \item the analysis of the turbulent-blob data (Fig.~\ref{fig:Matsuzawa}).
\end{itemize}
Claude also checked derivations, units and references, drafted revised passages of the text, including passages of the
sections comparing the theory with simulations and experiments, and assembled this paper from the author's earlier
manuscript. ChatGPT 5.5 (OpenAI) and Gemini 3.1 Pro (Google) were used to obtain critical reviews of a draft. These tools
and Claude Opus 4.6 were also used for copy-editing, exposition and \LaTeX\ formatting. The author directed the analyses,
decided the conclusions, reviewed the results and the text, and takes full responsibility for the content of the
manuscript. 

\section*{Supplementary Material}
The Supplementary Material for this article (the data processing referred to in the text as Supplementary Material, Sec.~S1) is provided as a separate file.

\section*{Data Availability Statement}
The data from~\cite{SreeniAkash2025} were quoted with the permission of the authors and are available on request.
The Max Planck wind-tunnel data of Ref.~\cite{GregXi2} were provided by the authors and are quoted with their permission; they are available from the authors. The code and data tables of the wind-tunnel comparison are available in Refs.~\cite{ZenodoNTDT3,LoopEquationsRepo}; the code of the DNS analyses of this paper will be added to Ref.~\cite{LoopEquationsRepo}. The turbulent-blob data of Ref.~\cite{Matsuzawa2026} are publicly available~\cite{MatsuzawaData2026}.
\section*{Acknowledgements}

We are grateful  to Prof. Sreenivasan  for valuable discussions and to Akash Rodhiya for providing the DNS data.

\bibliographystyle{Frontiers-Vancouver}
\bibliography{bibliography}

\end{document}

%% file: mymacros.tex
\def\8{\infty}

\def\undertext#1{\vtop{\hbox{#1}\kern 1pt \hrule}}

\def\VEV#1{\left\langle #1\right\rangle}

\def\bea{\begin{eqnarray} && &&}
\def\eea{\end{eqnarray}}

\let\oldexp\exp
\renewcommand{\exp}[1]{\oldexp\left(#1\right)}

\def\NS{Navier-Stokes}

\newcommand{\Mod}[1]{\ (\mathrm{mod}\ #1)}

\def\XXint#1#2#3{{\setbox0=\hbox{$#1{#2#3}{\int}$}
     \vcenter{\hbox{$#2#3$}}\kern-.5\wd0}}

\renewcommand{\Re}{\textbf{Re }}

\newcommand{\tpmod}[1]{{\@displayfalse\Mod{#1}}}
\usepackage[english]{babel}